\documentclass[useAMS,usenatbib]{mn2e}

\usepackage[dvipdfmx]{graphicx}
\usepackage{float}

\title[Simulating Submm galaxies]
{Sub-millimetre galaxies in cosmological hydrodynamic simulations: 
Source number counts and the spatial clustering} 
\author[Shimizu, Yoshida and Okamoto]{Ikkoh Shimizu,$^{1,2}$\thanks {E-mail:ikko.shimizu@ipmu.jp}
Naoki Yoshida$^{2,3}$ \thanks {E-mail:naoki.yoshida@ipmu.jp}, 
Takashi Okamoto$^{4}$ \thanks {E-mail:tokamoto@ccs.tsukuba.ac.jp} \\
$^{1}$College of General Education, Osaka Sangyo University, 3-1-1 Nakagaito, Daito, Osaka 574-8530, Japan \\
$^{2}$Kavli Institute for the Physics and Mathematics of the Universe, TODIAS, \\
The University of Tokyo, 5-1-5 Kashiwanoha, Kashiwa, Chiba 277-8583, Japan \\
$^{3}$Department of Physics, The University of Tokyo, 7-3-1 Hongo, Tokyo 113-0033, Japan \\
$^{4}$Center for Computational Sciences, University of Tsukuba, Tsukuba 305-8577, Japan}
\begin{document}

\date{In original form 2011 November 11}

\pagerange{\pageref{firstpage}--\pageref{lastpage}} \pubyear{2010}

\maketitle

\label{firstpage}

\begin{abstract}
We use large cosmological Smoothed-Particle-Hydrodynamics simulations
to study the formation and evolution of sub-millimetre galaxies (SMGs). 
In our previous work, we studied the statistical properties of ultra-violet
selected star-forming galaxies at high redshifts. 
We populate the same cosmological simulations with SMGs by calculating
the reprocess of stellar light by dust grains into far-infrared to millimetre
wavebands in a self-consistent manner.
We generate light-cone outputs
to compare directly the statistical properties of the simulated SMGs
with available observations.
Our model reproduces the submm source number counts 
and the clustering amplitude. 
We show that bright SMGs with flux $S > 1$ mJy reside in halos
with mass of $\sim 10^{13} M_{\odot}$ and have stellar masses greater than
$10^{11}~ \rm M_{\odot}$.
The angular cross-correlation between the SMGs and Lyman-$\alpha$ emitters
is significantly weaker than that between the SMGs and Lyman-break galaxies.
The cross-correlation is also weaker than the auto-correlation of the SMGs.
The redshift distribution of the SMGs shows a broad peak at $z \sim 2$,
where Bright SMGs contribute significantly to the global cosmic star formation 
rate density.
Our model predicts that there are hundreds of SMGs with $S > 0.1$ mJy 
at $z > 5$ per 1 square degree field. Such SMGs can be detected by ALMA.
\end{abstract}


\section{Introduction}
An array of recent millimetre and sub-millimetre (submm) observations 
revealed the physical properties of 
sub-millimetre bright galaxies (SMGs) 
with luminosities $10^{12} \sim 10^{13} {\rm L_{\odot}}$
\citep{Swinbank2004, Chapman2005, Capak2008, Coppin2009, 
Daddi2009, Tamura2009, Knudsen2010, Riechers2010, Hatsukade2011}. 
The power sources of such luminous SMGs are thought to be
very active star formation with $100-1000 \rm M_{\odot} yr^{-1}$ and/or
active galactic nuclei (AGNs).
Ultra-violet photons from massive stars or AGNs 
are converted to photons in submm to infrared bands 
by dust grains via their thermal emission. 
\citet{Alexander2003} find that,
in the Chandra Deep Field North survey, X-rays are detected from more than 
one-third of bright SMGs, which indicates a significant contribution from AGNs.

SMGs provide important information on the star formation rate, 
the stellar initial mass function, 
and the chemical evolution of the galaxies at high redshifts.
Because SMGs are highly obscured by dust, they often appear dark 
in UV and optical wavebands. 
Thus measuring their redshifts by conventional techniques is very difficult. 
Moreover, identifying the optical counterpart is not easy 
because the spatial resolution of the currently operating 
submm telescopes is much worse than that of large optical telescopes. 

So far, most of the SMGs with measured redshifts are at $z < 3$ 
\citep{Swinbank2004, Chapman2005}, 
but a few SMGs have been found at higher redshifts 
\citep{Capak2008, Coppin2009, Daddi2009, Knudsen2010}. 
SMGs generally show strong clustering 
\citep{Webb2003, Blain2004, Scott2006, Weis2009, Cooray2010, Hickox2012}. 
\citet{Maddox2010} measured the angular correlation of 
the $350~\rm \mu m$ and $500~\rm \mu m$ selected SMGs in the Herschel-ATLAS survey.
The SMGs are strongly clustered while the $250~\rm \mu m$ selected sample showed
weak or no clustering signals. 
It is thought that SMGs are very massive systems with large gas reservoir 
($\sim 10^{11}~\rm M_{\odot}$) 
and with large stellar masses ($\sim 10^{11}~\rm M_{\odot}$) 
\citep{Greve2005, Tacconi2006, Wardlow2011}. 
Also, SMGs are likely to be ancestors of massive elliptical galaxies 
in the local universe \citep{Lilly1996, Smail2004}. 

\citet{Chapman2005} report that the redshift distribution of 
SMGs shows a peak 
at $z \sim 2$, similarly to that of AGNs, 
although the size of the sample is small. 
Several SMGs have counterparts of star-forming galaxies such as
Lyman break galaxies (LBGs) and Lyman $\alpha$ emitters (LAEs)
\citep{Geach2005, Geach2007, Beelen2008, Daddi2009}.
However, many of high-redshift galaxies are not seen as SMGs 
probably because their submm fluxes are below typical observational 
flux limits ($> 1 ~\rm mJy$). 
Because observations so far detect very bright SMGs,
it remains unclear if SMGs are generally associated with LBGs/LAEs.
\citet{Matsuda2007} and \citet{Tamura2010} failed to detect submm sources 
at the location of LAEs in a protocluster region 
even though the target galaxies are very bright in Ly$\alpha$.

\citet{Dayal2010} use cosmological simulations 
and found that the submm fluxes of high-$z$ LAEs would be less 
than $0.1~ \rm mJy$.  
They argue, however, that many of such galaxies can be detected by ALMA. 
\citet{Yajima2011} study the evolution of submm flux of a simulated galaxy by 
using detailed three-dimensional radiative transfer calculations. 
They show that the simulated galaxy in a very early LAE phase can 
be detected by ALMA. 
Observing galaxies in the early formation phase is important 
to understand the early chemical evolution.

One of the most important observational quantities is the source number 
count of SMGs\citep{Greve2004, Laurent2005, Perera2008, Austermann2009, 
Austermann2010, Scott2010, Hatsukade2011}.
Unfortunately, there remain substantial 
uncertainties in theoretical models. 
\citet{Baugh2005} calculate the source number count of SMGs using a semi-analytic
galaxy formation model. 
Interestingly, in order to reproduce the observed source number count,
they make a crucial assumption that the stellar initial mass function (IMF) 
is 'top-heavy' when starburst occurs (see also \citet{Lacey2010}). 
More recently, \citet{Fontanot2007} 
reproduced the source number count of SMGs without 
assuming top-heavy IMF. 
They argue that the major difference between \citet{Baugh2005} and their study 
may be in the gas cooling model. 
In \citet{Fontanot2007}, the hot gas in a galaxy is assumed to cool more 
efficiently than in \citet{Baugh2005}. 
However, because of the efficient cooling, the model of \citet{Fontanot2007} 
overproduces 
massive galaxies at lower redshifts. 
It appears that some efficient star-formation recipe is
needed in the semi-analytic models, in order to reproduce the source number 
count of SMGs. It is interesting to see if the same is true 
for cosmological hydrodynamic simulations of galaxy formation. 
\citet{Dave2010} study the physical properties of SMGs using cosmological simulations.
They claim that smooth infalling gas or/and gas-rich satellites are 
important to produce large star formation rates. 
Although their finding seems reasonable, they employ a simple model that 
galaxies with very large 
star-formation rates are identified as SMGs.
They do not calculate dust absorption and the resulting submm flux.
It is important and timely to study in detail
the statistical properties of SMGs over a wide range
of redshift using cosmological hydrodynamic simulations 
by considering dust absorption and re-emission consistently.

In this paper, we study the statistical properties of SMGs such 
as the stellar mass function, 
the source number count and the angular correlation function.
To this end, we perform large cosmological hydrodynamic simulations 
based on the standard $\Lambda$CDM cosmology. 
Our simulations follow star formation, supernova feedback, 
and metal enrichment self-consistently. 
For the galaxies identified in our cosmological simulations,
we calculate the spectral evolution and dust extinction 
to estimate the FIR luminosity. 
In our earlier work \citep{Shimizu2011},
we used the same set of simulations to study the statistical properties
of LAEs at $z=3.1$. Our model successfully reproduced all the available 
observational data at $z=3.1$ such as the Lyman-$\alpha$ luminosity function,
the angular correlation functions, and the Lyman-$\alpha$ equivalent width
distribution. Our aim in the present paper is to build a consistent
theoretical model of galaxy formation
that reproduces the observed properties of both ultra-violet
selected galaxies and SMGs.

Throughout the present paper, we adopt the $\Lambda$CDM cosmology 
with the matter density $\Omega_{\rm{M}} = 0.27$, 
the cosmological constant $\Omega_{\Lambda} = 0.73$, 
the Hubble constant $h = 0.7$ in units of $H_0 = 100 {\rm ~km ~s^{-1} ~Mpc^{-1}}$ and 
the baryon density $\Omega_{\rm B} = 0.046$. 
The matter density fluctuations are normalised by setting
$\sigma_8 = 0.81$ \citep{WMAP}. 
All magnitudes are expressed in the AB system, 
and the Ly$\alpha$ EW$_{\rm Ly\alpha}$ values in this paper are 
in the rest frame. 

\section{Theoretical Model}
Our simulation code is based on an updated version of the Tree-PM 
smoothed particle hydrodynamics (SPH) code {\scriptsize GADGET-3} which 
is a successor of Tree-PM SPH code {\scriptsize GADGET-2} \citep{Gadget}.   
We employ $N = 2 \times 640^3$ particles in a comoving volume of 
100 $h^{-1}{\rm ~Mpc}$ on a side. 
The mass of a dark matter particle is 
$2.41 \times 10^8 h^{-1}{\rm M_{\odot}}$ 
and that of a gas particle 
is $4.95 \times 10^7 h^{-1}{\rm M_{\odot}}$, 
respectively. 
We implement relevant physical processes 
such as star formation, supernova feedback and 
chemical enrichment following \citet{Okamoto2008}, 
\citet{Okamoto2009} and \citet{Okamoto2010}.  
In \citet{Okamoto2010}, two wind models for supernova (SN) feedback are examined. 
One model assumes that the initial wind speed of gas around stellar particles that trigger SN 
is proportional to the local velocity dispersion of dark matter ($\sigma$). 
The model is motivated by recent observations of \citet{Martin2005}.
The other model assumes a constant initial wind speed, as has been used 
in many cosmological hydrodynamic simulations \citep{Springel2003, Nagamine2004}. 
The major difference between the two models is the wind mass-loading per star formation rate. 
In the former model, the mass-loading is proportional to $\sigma^{-2}$ whereas 
it is constant for all galaxies in the latter model. 
\citet{Okamoto2010} show that the former model reproduce the observed 
satellite luminosity 
function and the luminosity-metallicity relation in the Local Group satellites. 
We set the initial wind speed of gas
to be five times the local velocity dispersion, 
following \citet{Okamoto2010}.  

In our previous paper \citep{Shimizu2011}, 
we studied the statistical properties
of UV-selected galaxies.
We implemented wavelength dependent dust absorption to calculate the
UV and Lyman-$\alpha$ luminosities of the galaxies.
In the present paper, in order to calculate the FIR luminosity, 
we assume that the absorbed UV photons are
re-emitted in FIR by thermal dust grains.

\subsection{Dust absorption} 
The treatment of dust absorption is described in
\citet{Shimizu2011}. 
Briefly, we calculate the optical depth
$\tau_{\rm d}(\lambda)$ for UV continuum photons 
on the assumption that dust is distributed in a sphere
with a certain radius around a galaxy.
Then we use only one length scale, the effective radius
of dust distribution $r_{\rm d}$, to evaluate the optical depth. 
\begin{equation}
\tau_d(\lambda) = \frac{3 \Sigma_d Q(\lambda)}{4a_{\rm d} s},
\end{equation} 
where $a_{\rm d}$ and $s$ are the dust grain size
and the material density of dust grains, respectively,
and $\Sigma_d = M_{\rm d} / (\pi r_{\rm d}^2)$ is the dust surface mass density. 
For simplicity, we assume a single-sized dust component. 
We use the dust optical constant $Q(\lambda)$
as a function of wavelength given in \citet{Draine1984}, 
so that we obtain realistic SEDs for individual galaxies. 
We set $a_{\rm d} = 0.1 ~{\rm \mu m}$ and 
$s = 2.5 ~{\rm g}~{\rm cm}^{-3}$ for dust grains that 
are produced by supernovae \citep{Todini2001, Nozawa2003}. 
In this case, the resulting extinction curve is flatter 
than the SMC extinction curve and the Calzetti extinction curve, 
because $Q(\lambda)$ for large grains is nearly independent of wavelength
\citep{Hirashita2008, Hirashita2010}. 
Intriguingly, recent observations of ultraluminous infrared galaxies 
at $z \sim 1$ suggest that the extinction curve is 
flatter than the Calzetti extinction 
curve \citep{ShimizuT2011}. 
Furthermore, the extinction curve for the SN-produced 
dust size distribution reproduces the observation well
\citep{Hirashita2008, Hirashita2010}.

Although our single-sized dust model might seem crude, 
the resulting extinction curve does not significantly 
affect our main results presented in the paper. In section 4, 
we explicitly
show that adopting the steep SMC extinction curve 
yields virtually the same statistical quantities 
for the SMG population.

\begin{figure}
\includegraphics[width = 80mm]{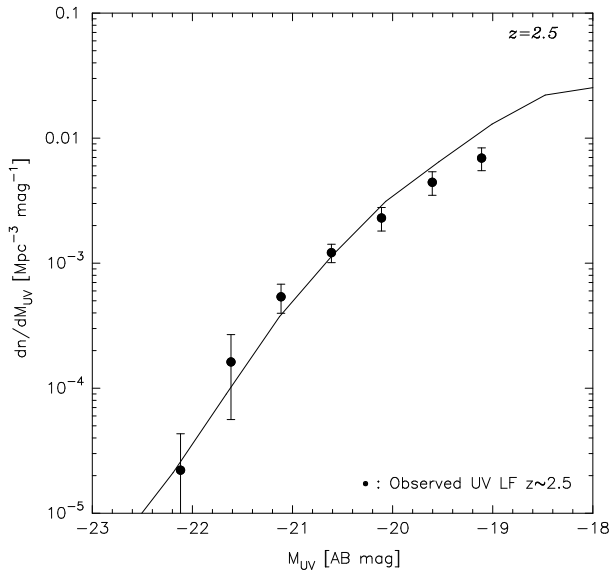}
\caption{UV ($1500~\rm \AA$ at rest-frame) luminosity function 
of the simulated galaxies at $z = 2.5$. 
The solid line shows our simulation result, which we compare with the 
observational data (points with error bars) of
\citet{Oesch2010}.
}
\label{UV_LF}
\end{figure}

We calculate the dust surface mass density as
$\Sigma_d = M_{\rm d} / (\pi r_{\rm d}^2)$
where $M_{\rm d}$ and $r_{\rm d}$ are the dust mass and 
the effective dust distribution radius of each simulated galaxy, respectively. 
The dust mass is calculated by 
\begin{equation}
M_{\rm d} = 0.01 m_{\rm g} \left( \frac{Z}{Z_{\rm \odot}} \right), 
\end{equation}
where $m_{\rm g}$, $Z$ and $Z_{\rm \odot}$ denote the gas mass, 
the gas metallicity 
and the solar metallicity, respectively \citep{Draine2007}. 
We adopt $Z_{\odot}=0.02$, and then the above equation 
gives $M_{\rm d} = 0.01 m_{\rm g}$ for $Z=Z_{\odot}$, which means 
that one percent (in mass) of the gas in a simulated galaxy with 
$Z=Z_{\odot}$ is locked in dust grains. Hence the dust mass scales
with the gas metallicity. 
We assume that the effective dust distribution radius of a galaxy 
$r_{\rm d}$ is a fraction of its virial radius, 
\begin{equation}
r_{\rm d} = e \times r_{\rm vir},
\end{equation}
where $e$ is a global parameter to be kept constant for all the galaxies.
The numerical value of $e$ is determined by matching the UV luminosity function
at $z = 2.5$ (see Fig. \ref{UV_LF}). The best fit value at $z=2.5$
is $e = 0.09$. 

We calculate the escape fraction of the UV continuum photons 
$f_{\rm c}$ using the following equation:
\begin{equation}
f_{\rm c}(\lambda) = \frac{1 - e^{- \tau_{\rm d}(\lambda)}}{\tau_{\rm d}(\lambda)}. 
\label{eq:escape_probability}
\end{equation}
The FIR luminosity $L_{\rm FIR}$ is then set equal to the total energy of 
UV photons absorbed by dust, 
\begin{equation}
L_{\rm FIR} = \int \left[ L^{\rm int}(\nu) - L^{\rm real}(\nu) \right] d\nu, 
\end{equation}
where $L^{\rm int}(\nu)$ is the intrinsic luminosity (per frequency) 
of the galaxy,
and $L_{\nu}^{\rm real}$ is the luminosity after dust absorption.
The above integration is evaluated over from UV to optical wavelengths. 
In the next subsection, we describe the calculation of the FIR flux. 

\subsection{Calculation of the FIR Flux}
\begin{figure*}
\includegraphics[width = 160mm]{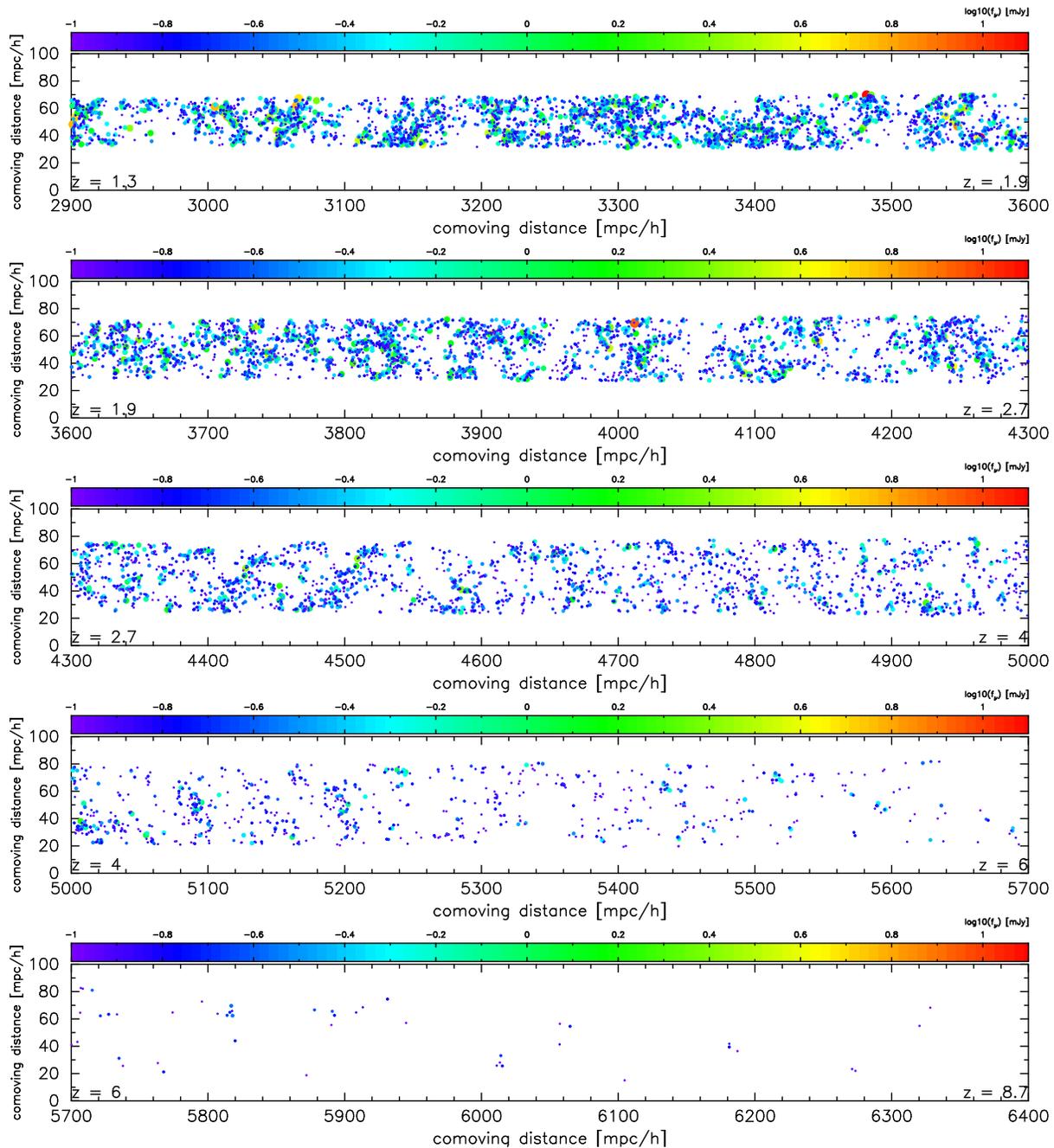}
\caption{The spatial distribution of our simulated SMGs on the past light 
cone of an observer.
We plot SMGs with submm flux greater than $0.1~\rm mJy$ at $270~\rm GHz$.
The redshift range is from $z = 1.3$ to $z = 8.5$. 
The field of view of the light cone is $40~\rm arcmin$ on a side. 
The colour indicates the submm flux; blue for faint SMGs and yellow 
to red for bright SMGs. 
Also the point size scales with the submm flux. 
The smallest points and the largest points correspond to 0.1 mJy 
and for 30 mJy, respectively. 
}
\label{SMG_LIGHT_CONE}
\end{figure*}
We calculate the FIR flux following \citet{Hirashita2002}. 
For the single-sized dust model, the monochromatic luminosity 
at frequency $\nu$ is written as
\begin{equation}
L_{\nu}^{\rm dust} = 4 \pi M_{\rm d} \kappa_{\nu} B_{\nu}(T_{\rm d}),
\label{DUST_FLUX}
\end{equation}
where $\kappa_{\nu}$, $B_{\nu}$ and $T_{\rm d}$ are 
the absorption coefficient, the Planck function and 
the dust grain temperature, respectively. 
The absorption coefficient $\kappa_{\nu}$ in FIR
is well described by a power-law $\kappa_{\nu} = \kappa_0 (\nu / \nu_0)^{\beta}$ 
with $\beta = 1 \sim 2$. 

Then the total FIR luminosity is 
\begin{eqnarray}
L_{\rm FIR} &=& \int_{0}^{\infty} L_{\nu}^{\rm dust} {\rm d}\nu \nonumber \\ 
            &=& 4 \pi M_{\rm d} \kappa_{\nu_0} \nu_0^{- \beta} \left( \frac{kT_{\rm d}}{h_{\rm p}} \right)^{4 + \beta} 
                \left( \frac{2h_{\rm p}}{c^2}\right)\nonumber \\
&\times& \int_0^{\infty} \frac{x^{3 + \beta}}{e^x - 1} dx, 
\label{eq:dust_luminosity}                
\end{eqnarray}
where $k$, $h_{\rm p}$ and $c$ are the Boltzmann constant, the Planck constant, 
and the speed of light, respectively. We denote $x = h_{\rm p} \nu / (kT_{\rm d})$. 
We set a normalization parameter $\nu_0 = 3.0 \times 10^{12} \rm Hz$ 
which corresponds to the wavelength of $100 {\rm \mu m}$. 
The absorption coefficient $\kappa_0$ is given by
$\kappa_0 = 3 Q_{\nu_0} (4as)^{-1}$, 
where $Q_{\nu_0}$ is the dust optical constant at $\nu_0$. 
The value of $\kappa_0$ is $46.95~ \rm g^{-1} cm^{2}$. 
Note that $Q_{\nu_0}/a$, and hence $\kappa_0$, are independent of dust size.

Using equation (\ref{eq:dust_luminosity}), we derive 
the dust temperature $T_{\rm d}$ as 
\begin{equation}
T_{\rm d} = 7.64 \left( \frac{L_{\rm FIR} / L_{\odot}}{M_{\rm d} 
/ {\rm M_{\odot}}}\right)^{1/6} [\rm K], 
\end{equation}
where we set $\beta = 2$ assuming graphite and silicate dust grains 
\citep{Draine1984}. 
Finally, the observed submm flux of a galaxy is obtained from
\begin{equation}
f_{\nu}^{\rm obs} = \frac{(1 + z) L_{\nu (1 + z)}}{4\pi d_{\rm L}^2}, 
\end{equation}
where $d_{\rm L}$ is the luminosity distance to the galaxy. 
The dust temperature for our sample galaxies is typically 
$T_{\rm d} = 33~ \rm K$. 
To directly compare our model with the observational results of the 
AzTEC observation, 
we consider the filter response function of the AzTEC detector 
when we calculate the submm flux at the observed frame $1.1~ \rm mm$.

\section{Result}

\subsection{UV luminosity function}
We first examine whether our model reproduces the UV luminosity function.
This is an important check because the total FIR luminosity is equal 
to, by definition in our model, the total UV luminosity absorbed by dust grains. 
Fig. \ref{UV_LF} shows the UV ($1500~\rm \AA$ at rest-frame) luminosity function 
of our model at $z = 2.5$. 
Clearly, our model reproduces the observational data very well. 
Given the equally good agreement between our model predictions
for $z=3.1$ LAEs \citep{Shimizu2011} and other available 
observational data of UV-selected galaxies (Shimizu et al., in preparation),
we are confident that our cosmological simulations can be used
to build a self-consistent model for UV-selected galaxies and SMGs
at high redshifts.

\subsection{The light cone output}
Most of the recent observations do not provide redshifts for SMGs.
In order to directly compare our model with the observations,
we generate a light cone output which extends 
from $z = 1.3$ to $z = 8.5$ using a number of simulation outputs. 
We coordinate the output redshifts
to fill the volume of a lightcone from $z = 1.3$ to $z = 8.5$ without gaps.
We then randomly shift each simulation box so that the same objects 
(at different epochs) do not appear multiple times on a single line-of-sight. 
Fig. \ref{SMG_LIGHT_CONE} shows the distribution of simulated SMGs 
on the light cone from $z = 1.3$ to $z = 8.7$. 
For this figure, we set the SMG detection limit to be $0.1~\rm mJy$ at $270~\rm GHz$.
Most of the SMGs are located at $z = 1-4$,
but there are dozens of SMGs even at $z > 6$. 
Such faint SMGs can be detected by ALMA with less than one 
hour integration using 16 antennas at the $350 ~\rm GHz$ band. 
The high redshift SMGs will provide invaluable information
on star formation activities in the early universe.

\subsection{The source number counts}
 
\begin{figure}
\includegraphics[width = 80mm]{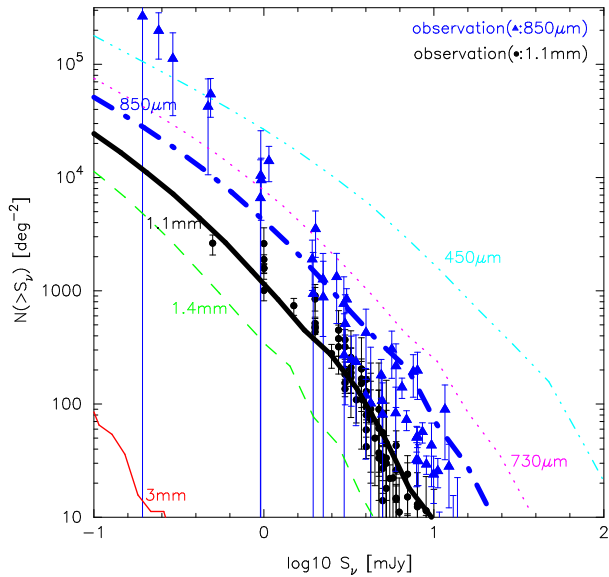}
\caption{The source number counts of SMGs for six observed frame fluxes. 
The thick solid line shows the number count 
at the observed frame $1.1~\rm mm$ (AzTEC detector band). 
The thick dash-dotted line is for the observed frame $850~\rm \mu m$,
and the other lines are for
the observed frames $3.0~ \rm mm$, $1.4~ \rm mm$, $730~ \rm \mu m$ 
and $450~ \rm \mu m$ 
which correspond the five ALMA bands, respectively.  
The filled circles with error bars show the observed number counts 
at $1.1~ \rm mm$ by AzTEC  
\citep{Greve2004, Laurent2005, Perera2008, Austermann2009, Austermann2010, Scott2010, Hatsukade2011}. 
The filled triangles with error bars show the observed number counts at $850~\mu m$ \citep{Chapman2005}.
}
\label{NUM_COUNT}
\end{figure}

\begin{figure}
\includegraphics[width = 80mm]{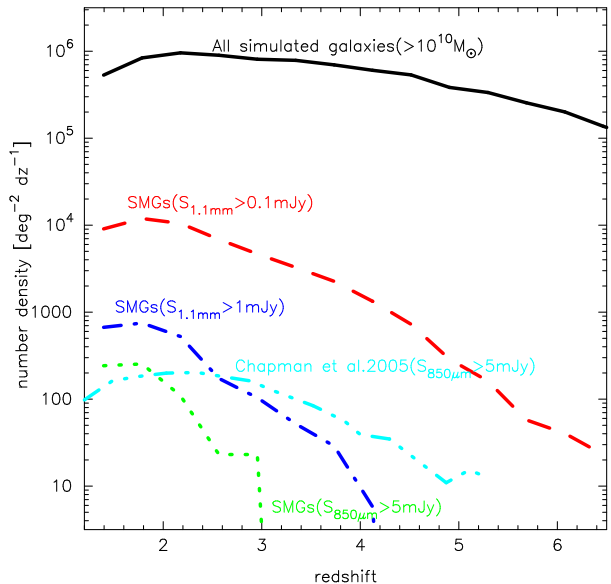}
\caption{The evolution of the number density of SMGs as a function of redshift. 
The dashed and dash-dotted lines represent our model prediction 
for the detection limit of $0.1~ \rm mJy$ and $1~ \rm mJy$ 
at the observed frame $1.1 \rm mm$, respectively. 
The dotted line is also our model prediction for a high detection 
limit of $5~\rm mJy$ at the observed frame $850~\rm \mu m$,
to be compared with the observation of \citet{Chapman2005}.
For reference, we plot the number density of all simulated galaxies (solid line) 
with halo masses greater than $10^{10}~\rm M_{\odot}$. }
\label{NUM_COUNT_EVILUTION}
\end{figure}

\begin{figure}
\includegraphics[width = 80mm]{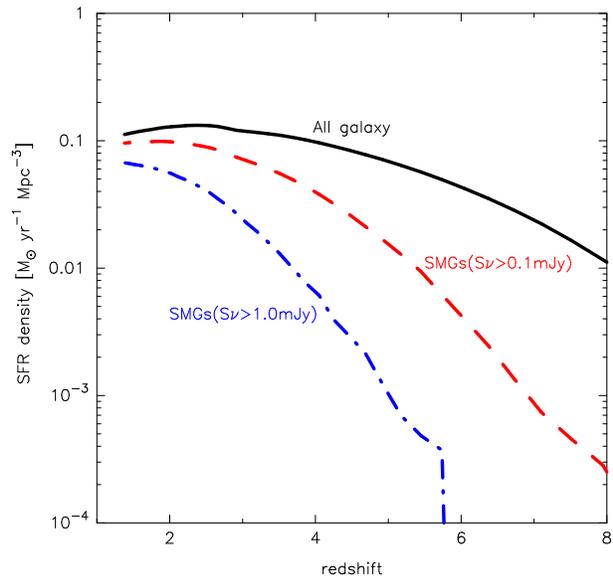}
\caption{The star formation rate density contributed by SMGs.
The solid, dashed and dash-dotted lines are for
all the galaxies,
SMGs with $S > 0.1~ \rm mJy$ at observed frame $1.1~\rm mm$
and SMGs with $S > 1~ \rm mJy$ at observed frame $1.1~\rm mm$,
respectively.}
\label{SMG_SFH}
\end{figure}

We use the generated light cone output to conduct mock observations 
of the SMGs.
Fig. \ref{NUM_COUNT} shows the number counts of SMGs. 
The filled circles and the filled triangles are 
the observational results at the observed frame 
$1.1~\rm mm$ and $850~\rm \mu m$, respectively. 
The lines shows our model predictions for various
bands.
To directly compare our model with the observation, 
we use the filter response function of the AzTEC detector 
when we calculate the submm flux at the observed frame $1.1~ \rm mm$. 

Our model reproduces the observational results remarkably well.
The $850~\rm \mu m$ source count is somewhat smaller than the observation at
$S < 1$ mJy, but the error bars are also large at the faint end. 
The thin solid, dashes, dotted and double-dotted lines are the 
predicted SMG number counts
at observed frames $3.0~ \rm mm$, $1.4~ \rm mm$, 
$730~ \rm \mu m$ and $450~ \rm \mu m$,
which correspond to the ALMA detector bands. 
The number of SMGs increases with shorter wavelength,
suggesting that SMGs are detected easier at shorter wavelengths; 
this is another manifestation of the negative k-correction 
for SMGs.

\subsection{The SMG number density evolution}
In this subsection, we study the redshift evolution of the SMG population
in more detail.
Fig. \ref{NUM_COUNT_EVILUTION} shows the evolution of the number density of SMGs. 
The dashed and dash-dotted lines represent the number counts for the
detection limit of $0.1~ \rm mJy$ and $1~ \rm mJy$ at the observed frame $1.1~ \rm mm$. 
We also plot our model number count 
at $850~\rm \mu m$ with the detection limit $5~\rm mJy$. 
The triple-dotted line shows the observational result at 
$850~\rm \mu m$ from \citet{Chapman2005}. 
Note that the sample of \citet{Chapman2005} 
consists of SMGs in multiple fields with various
areas. Thus we normalize the observed redshift distribution
arbitrarily to make the comparison easy. 
We also plot the number of all the simulated galaxies (the solid line) 
with halo mass more than $10^{10}~\rm M_{\odot}$. 
Regardless of the selection, the SMG number counts peak at $z \sim 2-3$. 
Our model lacks very bright SMGs at high redshifts. 
This is likely due to the small simulation box size (cosmic variance). 
Note also that, in general, determining the redshift distribution of the 
SMGs are not trivial if only submm observations are used.

In our model, approximately $90\%$ of SMGs with $> 0.1~{\rm mJy}$ 
at $1.1~{\rm mm}$ are at $z>2$.
This implies that most of the observed SMGs are high-$z$ galaxies, 
and the contribution from low-$z$ galaxies is small. 

We calculate the contribution from SMGs to the global cosmic star 
formation rate (SFR) evolution.
Fig.  \ref{SMG_SFH} shows the
sum of the SFRs of all the simulated galaxies (solid line),
that of SMGs with $> 0.1~ \rm mJy$ at observed frame $1.1~\rm mm$ (dashed line)
and that of SMGs with $> 1~ \rm mJy$ at observed frame $1.1~\rm
mm$ (dot-dashed line).

The contribution of SMGs to the cosmic SFR is small at
$z > 6$.
However the contribution rapidly increases with decreasing redshift.
At $z \sim 2$, Bright SMGs with $S > 1~ \rm mJy$ at $1.1~\rm mm$
contributes roughly one third to the total SFR.
Clearly, it is important to include SMGs to estimate the
true global star-formation rate at $z \sim 1 - 4$.

\begin{figure*}
\includegraphics[width = 160mm]{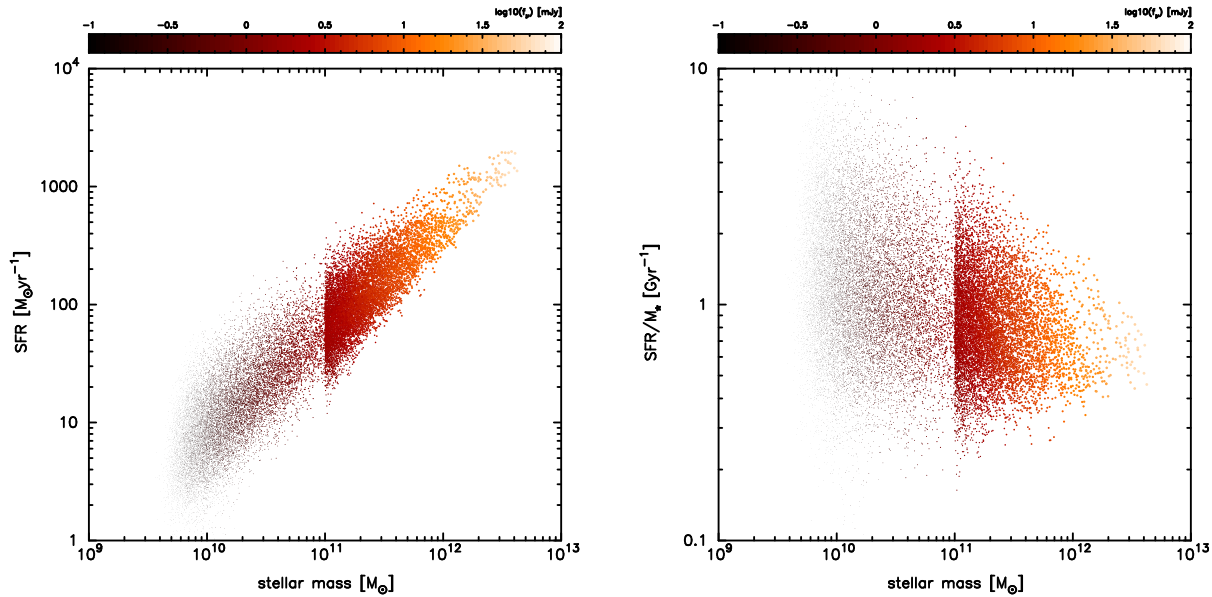}
\caption{The star formation rates of the simulated SMGs 
(left) and the specific star formation rates (right) as a function of the stellar mass.
We use bigger size points for brighter SMGs. Also the colour indicates the submm flux.
To reduce the number of points, we randomly chose one tenth of the galaxies
for stellar masses below $10^{11} M_{\odot}$.}
\label{SMG_STELLAR_MASS}
\end{figure*}

\subsection{The stellar mass of SMGs}
Fig. \ref{SMG_STELLAR_MASS} shows the SFR (the left panel) and specific SFR (the right panel) 
as a function of the stellar mass. 
We plot all the SMGs in our light cone data from $z = 1.3$ to $z = 8.7$. 
There are tens of SMGs with the SFR greater than $1000~ \rm M_{\odot} yr^{-1}$,
whereas the specific SFR is roughly constant ($\sim 10^{-9}~ \rm yr^{-1}$). 
Bright SMGs have very large stellar masses in excess of $10^{12}~\rm M_{\odot}$.
Some of our simulated SMGs with largest stellar
masses ($ > 10^{12}~\rm M_{\odot}$) consist of multiple clumps within 
a single host halo. 
However, we do not treat each such clump
of star particles as an individual galaxy. 
Essentially we regard one halo as a SMG.
This crude definition is motivated by the fact that the current observations do not
have high enough angular resolution to resolve the internal structure within a halo.
See also the discussion in section 3.6 and Fig.  \ref{SMG_MASS}
which shows the host halo mass of the SMGs.

\subsection{The distribution of SMGs and the angular correlation functions}
\begin{figure}
\includegraphics[width = 80mm]{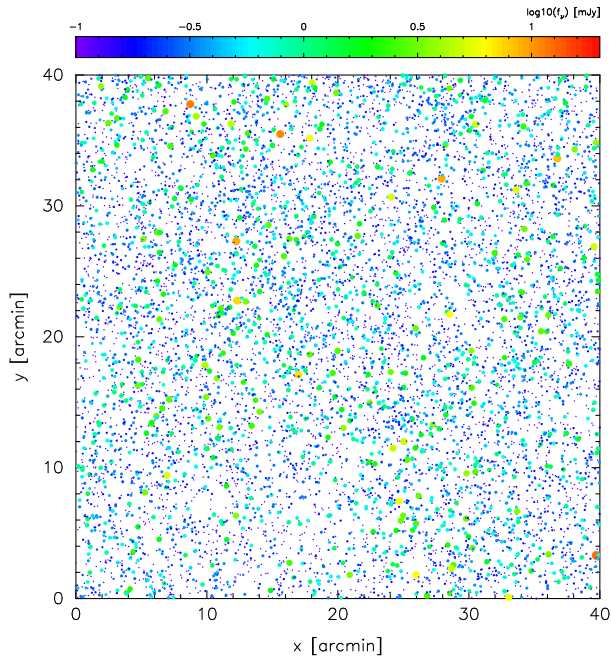}
\caption{The projected distribution of SMGs in a field-of-view 
of 40 arcmin $\times$ 40 arcmin. 
The left and right panel show SMGs with $S > 0.1~ \rm mJy$.  
The point size scales with the submm flux. The smallest points and the largest points 
are for $\sim 0.1$ mJy and for $\sim 30$ mJy, respectively.
Also the point colour indicates the submm flux.}
\label{SMG_DIST}
\end{figure}
Fig. \ref{SMG_DIST} shows the projected distribution of the
SMGs on the lightcone. 
We plot the SMGs with $S > 0.1~ \rm mJy$ at the observed frame $1.1~ \rm mm$. 
As is naively expected, brighter SMGs are strongly clustered,
whereas faint SMGs appear to be distributed more or less homogeneously
over the field of view.

\begin{figure}
\includegraphics[width = 80mm]{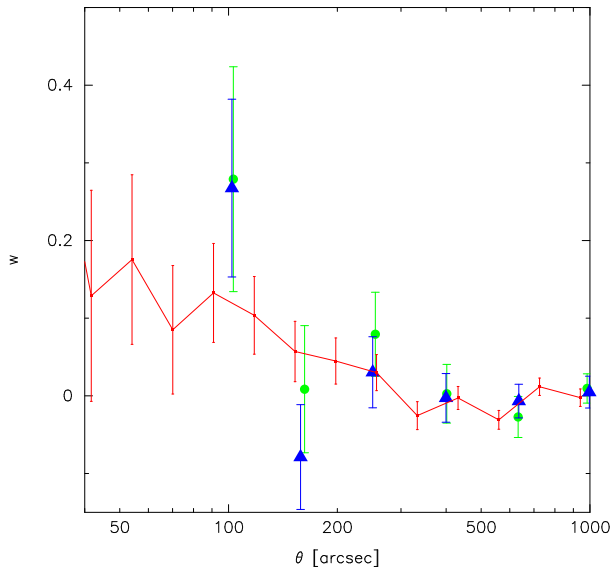}
\caption{We plot the angular two-point correlation function of the simulated SMGs
(solid line with error bars).
The points with error bars are the observational data of \citet{Williams2011}.  
Different symbols refer to different catalogues of SMGs.
The circle and triangle points are $3.5\sigma$ sources and $3\sigma$ sources, respectively. }
\label{ACF}
\end{figure}

Next, we calculate the angular two-point correlation function (ACF) of the SMGs 
to quantify the clustering strength. 
In Fig. \ref{ACF}, 
the solid line is the correlation function of our simulated SMG sample.
For comparison, we have chosen the same detection 
limit of $1.26~ \rm mJy$ as the observation of \citet{Williams2011}.
Our simulation result is consistent with the observational data although 
with substantial scatters for both the data.

We calculate the angular cross correlation between the simulated SMGs 
and other UV-selected galaxies. 
We use one simulation box at $z = 3.1$ for this calculation.
We consider two flux limits for our simulated SMGs,
$0.1~ \rm mJy$ and $1~ \rm mJy$ at observed frame $1.1~\rm mm$. 
The number of simulated SMGs for the two cases is then $993$ and $18$, respectively. 
We identify simulated galaxies which satisfy ${\rm EW_{Ly\alpha} > 30~{\rm \AA}}$ 
and $L_{\rm Ly\alpha}^{\rm obs} > 1.0 \times 10^{42} {\rm [erg / s]}$ as LAEs \citep{Shimizu2011}. 
Here $L_{\rm Ly\alpha}^{\rm obs}$ and $\rm EW_{Ly\alpha}$ are the observed Ly$\alpha$ luminosity 
and the rest frame Ly$\alpha$ equivalent width. 
We define LBGs at $z \sim 3$ by the following selection: 
\begin{eqnarray}
23 < & z_{\rm AB} & \leq 26.0, \\
U_{\rm AB} - V_{\rm AB} \geq 0.8, && V_{\rm AB} - z_{\rm AB} \leq 2.7, \\
U_{\rm AB} - V_{\rm AB} \geq & 1.8 &(V_{\rm AB} - z_{\rm AB}) + 1.6,
\end{eqnarray}
where $U_{\rm AB}$, $V_{\rm AB}$ and $z_{\rm AB}$ denote AB magnitude of U band, V band and z band, respectively. 
We select the LBGs in this way consistently with the observation of  \citet{Yoshida2008}. 
Finally, we locate $2913$ LAEs and $3211$ 
LBGs in the simulation box at $z\sim3$.

\begin{figure*}
\includegraphics[width = 160mm]{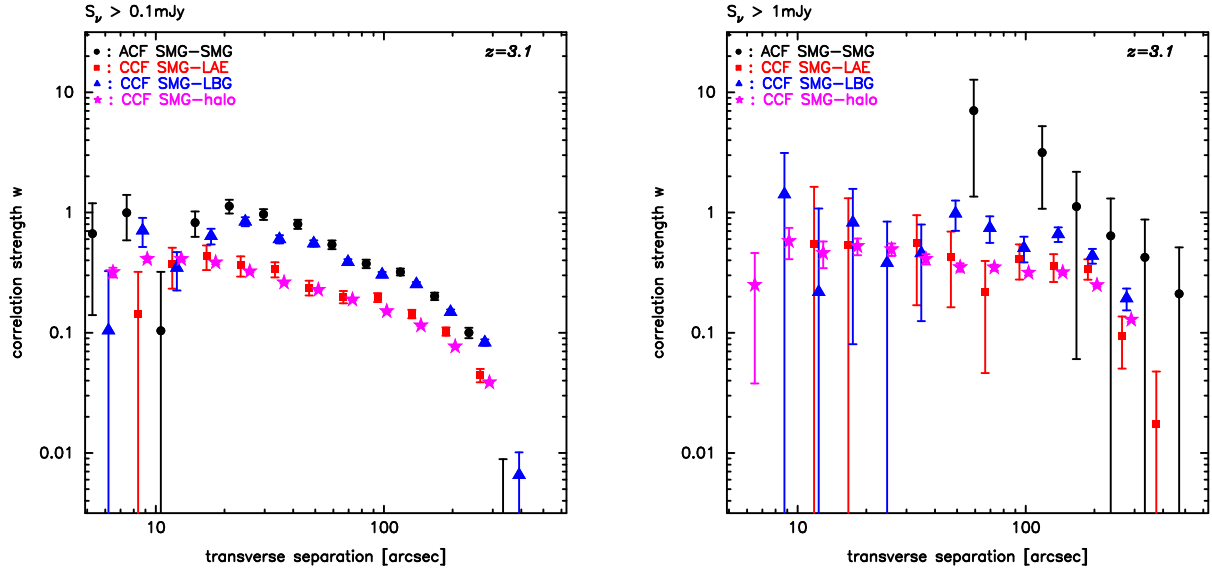}
\caption{
We plot the angular cross-correlation with two other populations of galaxies;
the SMG-LAE cross-correlation (squares with error bars)
and the SMG-LBG cross-correlation (triangles with error bars). 
For comparison, we also plot the angular auto-correlation of the SMGs (solid circles with error bars) 
and the angular cross correlation between the SMGs and dark matter halos 
with $10^{10}~\rm M_{\odot}$ (stars with error bars). 
We set two detection limits for the submm flux; $0.1~ \rm mJy$ for the left panel 
and $1~ \rm mJy$ for the right panel, respectively.}
\label{CCF}
\end{figure*}

Fig.  \ref{CCF} shows 
the SMG-LAE and SMG-LBG angular cross-correlation functions.
For comparison, we also plot the auto correlation function of SMGs (solid circles with error bars) 
and the cross-correlation between SMGs and dark halos 
with mass greater than $> 10^{10}{\rm M_{\odot}}$ (stars with error bars).  
For the SMG flux limit of $0.1~ \rm mJy$,
the SMG-LBG cross correlation is significantly stronger than the SMG-LAE cross correlation. 
The former is indeed very similar in the shape and the amplitude to the auto-correlation of
the SMGs, implying that the SMGs are a similar population to the LBGs (see also the 
SFR shown in Fig.  \ref{SMG_STELLAR_MASS}).
We have checked that many UV bright galaxies in our sample are 
identified as SMGs.

\subsection{The host halo mass of SMGs}
Fig. \ref{SMG_MASS} shows the submm flux of simulated SMGs as a function of their host halo mass.
We use all the SMGs in our light cone data from $z = 1.3$ to $z = 8.7$. 
The submm flux of simulated SMGs roughly scales with their host halo mass, 
but with large dispersions for low mass objects. 
We show the dispersion of host halo mass and submm flux in each mass bin. 
We also plot dark matter halo mass function at $z = 2$.  
Clearly, bright SMGs with ($S > 1~ \rm mJy$) reside in rare, large halos with
several times $10^{12}~ \rm M_{\odot}$. 

It is important to note that, because the angular resolution of the 
current mm/submm observations 
is still limited to $\sim 30-60$ arcseconds, multiple galaxies
in a galaxy-group size halo are not well resolved. Future observations
using ALMA will likely find multiple sources in massive halos. 
>From Fig. \ref{SMG_MASS}, we expect that
SMGs hosted by smaller halos
with several times $10^{11}~ \rm M_{\odot}$ can be detected
by observation with a detection limit of $0.1~ \rm mJy$. 

\section{Conclusions and Discussion}
We have studied a variety of statistical properties of SMGs using a 
large cosmological hydrodynamic simulation. 
The simulation follows the formation and evolution of star-forming galaxies 
by employing a new feedback model of \citet{Okamoto2010}. 
Earlier  in \citet{Shimizu2011}, the same simulation was used to study
the properties of LAEs at $z = 3.1$.
We have shown in the present paper that our SMG model is consistent with 
a number of observational data currently available.
The redshift distribution of the SMGs peaks broadly at $z \sim 2$. 
Most of the SMGs are at $z > 2$, and  
the contribution from low-$z$ ($z < 1$) galaxies to the submm 
source count is small.
The contribution from the SMGs to the global star formation rate density 
increases with decreasing redshift. Bright SMGs contributes significantly
to the star formation rate at $z \sim 2-3$. 
We calculate the angular two point correlation to quantify the clustering 
amplitude of SMGs. 
Our result is consistent with the observations of \citet{Williams2011}. 
We also study the angular cross correlation functions between 
SMGs and LBGs and the correlation between SMGs and LAEs. 
The former is significantly
stronger than the latter. 
Considering the high star-formation rates of the simulated SMGs, 
we argue that SMGs are similar population to LBGs. 
Finally, we explicitly showed 
that the bright SMGs preferentially reside in massive halos ($> 10^{12}~\rm M_{\odot}$) 
and that their typical stellar mass are greater than $10^{11}~ \rm M_{\rm \odot}$. 


In our simple model of a single-sized, supernovae produced dust, 
the resulting extinction curve is flatter than the standard ones
such the SMC extinction curve or Calzetti extinction curve. 
This mean that the reprocess of UV photons to FIR by dust 
works more efficiently in our model. However, interestingly, we have found 
that the characteristic
feature of the extinction curve does not significantly
affect our main results.

\begin{figure}
\includegraphics[width = 80mm]{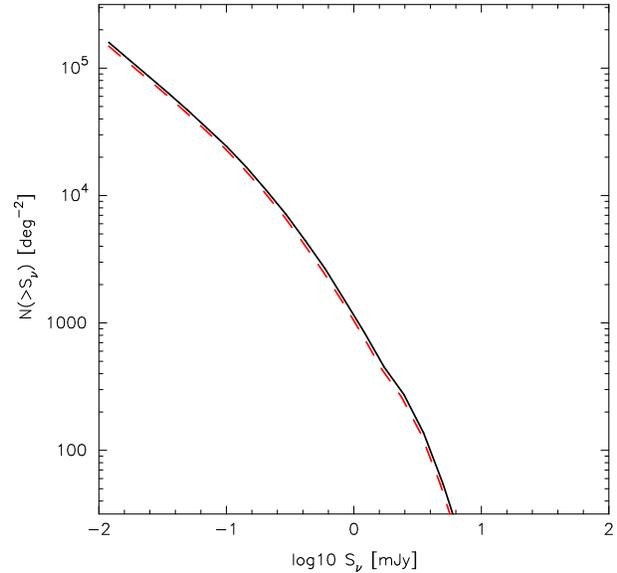}
\caption{SMG number count at 1.1 mm for the two dust extinction models; 
our model based on a single-sized dust (solid) and 
the other using the SMC extinction curve (dashed).}
\label{SMG_NC_COMPARISON}
\end{figure}

\begin{figure}
\includegraphics[width = 80mm]{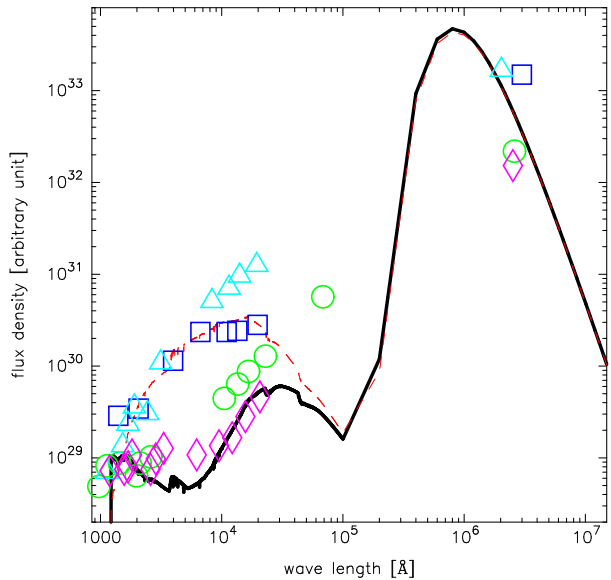}
\caption{An example SED of one of our simulated SMGs. 
The thick solid line and thin dashed line represent the dust-extinct 
SED for our simple dust model 
and that for the SMC extinction curve, respectively. 
We also show the observed SEDs for several SMGs \citep{Michalowski2010}
by open circles, open squares, open triangles and open diamonds.
The luminosity of the observed galaxies are arbitrary normalized for 
this comparison.   
}
\label{SMG_SED}
\end{figure}

For direct comparison, we calculate the FIR luminosities
for our galaxy samples using the SMC extinction curve. 
We then estimate the SMG number count following otherwise the same procedures as in
our fiducial model. 
Fig. \ref{SMG_NC_COMPARISON} compares the source number count of our model (solid line) 
and the one calculated using the SMC extinction curve (dashed line). 
Clearly the difference between our base model and the other is very small. 
This can be explained by the fact that we normalize the overall luminosity 
of the galaxies by matching the UV luminosity 
function at rest-frame $1500~{\rm \AA}$ to the observed one (see Fig. \ref{UV_LF}).
Even for a different extinction model, the overall level of extinction 
in UV is always normalized at $1500~{\rm \AA}$.
Therefore, the difference in the exact shape of the extinction curves near UV does not affect 
significantly the SMGs number count in our model, as is explicitly shown in Fig. \ref{SMG_NC_COMPARISON}.
Note, however, that the UV to optical colour of individual galaxies is visibly affected by the extinction
curve. Our base model predicts that simulated galaxies appear bluer in UV to optical range
than in the case with SMC or Calzetti extinction curve models. 
Fig. \ref{SMG_SED} shows, as an illustrative example, 
the SED of one of our SMG samples
from UV to millimetre.
The SED for the same galaxy but with SMC extinction curve
is also shown.
We compare them with the observed SEDs for several 
SMGs \citep{Michalowski2010}. 
Some SMGs have flat (blue) SEDs at UV to optical range whereas others 
appear redder. The full SEDs are measured only for a limited
number of samples. It would be highly interesting to
observe SMGs in detail in rest frame optical to UV 
in order to derive the dust extinction law,
and hence the dust size distribution, for star burst galaxies. 

\begin{figure}
\includegraphics[width = 80mm]{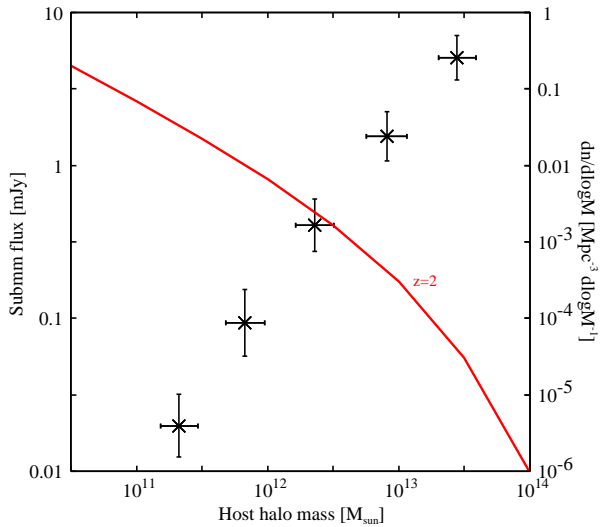}
\caption{The mean submm flux of our simulated SMG samples as a function of their host halo mass. 
We also plot the halo mass function at $z = 2$ (the solid line).
Bright SMGs with $S > 1$ mJy are hosted by dark halos with mass greater
than $\sim 10^{13} M_{\odot}$. }
\label{SMG_MASS}
\end{figure}

Next, we discuss the source confusion limit of submm observations. 
We have seen in previous sections (Fig. \ref{SMG_STELLAR_MASS} and Fig. \ref{SMG_MASS}), 
that some simulated SMGs have large host halo masses with $> 10^{13} \rm M_{\odot}$ 
and/or large stellar masses with $> 10^{12} \rm M_{\odot}$. 
Such massive systems often have multiple substructures (main and satellite galaxies). 
Basically each such satellite galaxy should be treated as an independent galaxy.
Note, however, that the angular resolution of the current submm/mm observation is worse than 
that of optical observation. 
For example, the resolution of the AzTEC telescope is $30$ arcseconds 
which corresponds $260~ \rm kpc~ (physical~ scale)$ at $z = 2$. 
The virial radii of the galactic halos in our simulation 
are only as large as $\sim 200 ~\rm kpc~ (physical~ size)$ at $z = 2$.
Thus, our treatment that a halo hosts one SMG as a whole is reasonable,
even realistic, if we compare with observations with angular resolution 
of 0.5-1 arcminutes. 

 Finally, we discuss the evolution of the stellar mass function.
Our galaxy formation model is defined at high redshift, rather than being
calibrated against galaxies at the present epoch.
We choose a few model parameters such as the overall
normalization of dust extinction to reproduce the observational
data of star-forming galaxies at high-$z$ ($z \sim 3$)\citep{Shimizu2011}.
Contrastingly, previous semi-analytic models (e.g., 
\citet{Baugh2005} and \citet{Lacey2010})
determined the main physical parameters to match the observations
of the present-day galaxies. For example, 
\citet{Baugh2005} needed to adopt an extreme IMF (top-heavy IMF) 
for starburst galaxies at high redshift to reproduce the observed SMG number
counts.
More recently, \citet{Fontanot2007} reproduced the SMGs number count
adopting the Salpeter IMF for all their galaxy samples.
However, their model assumes more efficient gas cooling in galactic haloes 
than in
\citet{Baugh2005} and \citet{Lacey2010}.
Consequently, very massive galaxies with high star-formation rates are
formed, which results in disagreement in the stellar mass function at low
redshifts ($z < 1$).
Clearly it is important to study the stellar mass function of our model.
\begin{figure}
\includegraphics[width = 80mm]{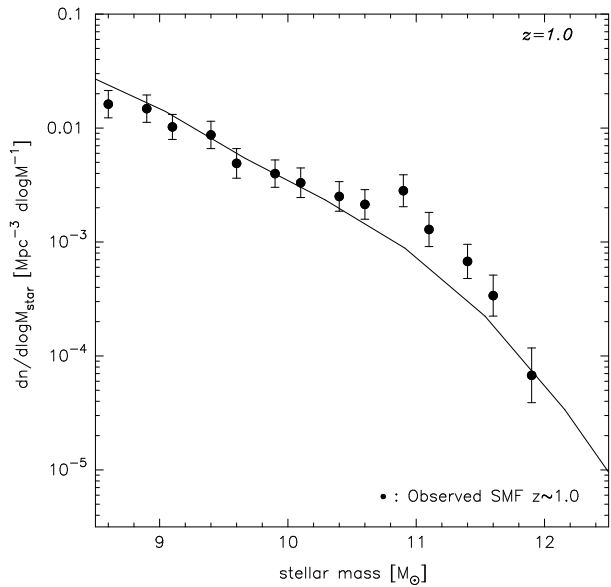}
\caption{The stellar mass function at $z = 1$. The solid line is
our simulation result.
The points with error bars show the observational result
from the GOODS NICMOS survey \citep{Mortlock2011}.}
\label{SMF}
\end{figure}

Fig. \ref{SMF} shows the stellar mass function of our simulated galaxies at
$z = 1$.
We also plot the observational data (solid points with error bars) of
\citet{Mortlock2011}.
Our model is in good agreement with the data from the GOODS NICMOS survey 
as is clearly seen in Fig. \ref{SMF}. This provides yet another
support for our galaxy formation model that covers a multiple 
populations, from LAEs, LBGs to SMGs.

In order to study the present-day stellar mass function, 
we have run our cosmological simulation down to $z=0$.
Our simulation over-predicts the number density of massive galaxies
with stellar mass greater than $10^{11} M_{\odot}$. 
Therefore, although we have successfully constructed a consistent model
for SMGs and UV-selected galaxies at high redshifts, 
it appears that our galaxy 
formation model lacks some physical mechanism(s) that shapes the present-day
stellar mass function. As an attempt, we have implemented a model
where gas cooling is quenched in large galaxies whose
halos' velocity dispersions are greater than 141 km/sec
\footnote{We have done a few test calculations by varying the
threshold velocity dispersion and have found that the run
with $\sigma_{\rm th} = 141$ km/sec reproduces the
break of the observed $z=0$ stellar mass function at
$M_{*}\sim 10^{11} M_{\odot}$.}.
Then the resulting stellar mass function closely agrees with
the observational data at $z=0$ derived by Li \& White (2009).
We argue that the kind of strong feedback, in terms of
star formation efficiency, may be needed 
for viable galaxy formation models, in order for them
to reproduce all the available data from low to high redshifts.
Constructing such a perfect model is the ultimate goal
of the study of galaxy formation, but it is beyond the scope
of the present paper. Further studies on the efficiency
of star formation in galaxies are clearly needed using 
multiple approaches.

Together with our previous study on Lyman-$\alpha$ emitters 
at $z=3.1$\citep{Shimizu2011},
we now provide a unified galaxy 
population model within the standard $\Lambda$ Cold Dark Matter
cosmology, which reproduces simultaneously the statistical properties
of UV-selected star-forming galaxies and sub-millimetre galaxies
at high redshifts.

\section*{Acknowledgments}
The authors are grateful to B. Hatsukade for providing 
their observational data. 
IS would like to thank B. Hatsukade, Y. Tamura, T. Nozawa, K. Nagamine 
and M. Kobayashi for stimulating discussion. 
Numerical simulations have been performed with the EUP, PRIMO 
and SGI cluster system installed 
at the Institute for the Physics and Mathematics of the Universe, 
University of Tokyo. 
This work is partially supported by Grant-in-Aid for Young 
Scientists (S) (20674003)
and by the FIRST program Subaru Measurements of Images and Redshifts (SuMIRe)
by the Council for Science and Technology Policy.
TO acknowledges the financial support  of Grant-in-Aid for Young Scientists
(B: 24740112) and by MEXT HPCI STRATEGIC PROGRAM.

\bsp

\label{lastpage}

\end{document}